\documentclass[twocolumn,prc,aps,eqsecnum,amssymb,floatfix,showpacs]{revtex4}
\usepackage{bm}
\usepackage{graphicx}

\begin{document}
%
\title{Muon capture on deuteron and $^3$He}

\author{L.E. Marcucci$^{1,2}$, M. Piarulli$^3$, M. Viviani$^2$, 
L. Girlanda$^{1,2}$, A. Kievsky$^2$, S. Rosati$^{1,2}$, and R. Schiavilla$^{3,4}$}

\affiliation{
$^1$ \mbox{Department of Physics, University of Pisa, 
56127, Pisa, Italy} \\
$^2$  \mbox{INFN-Pisa, 56127, Pisa, Italy} \\ 
$^3$  \mbox{Department of Physics, Old Dominion University, 
Norfolk, VA 23529, USA} \\  
$^4$  \mbox{Jefferson Lab, Newport News, VA 23606, USA} \\}

\begin{abstract}
The muon capture reactions $^2$H($\mu^-,\nu_\mu$)$nn$ and 
$^3$He($\mu^-,\nu_\mu$)$^3$H are studied with 
conventional or chiral realistic potentials
and consistent weak currents. The initial and final $A=2$ and 3 
nuclear wave functions are obtained from the Argonne $v_{18}$ or 
chiral N3LO two-nucleon potential, 
in combination with, respectively, the Urbana IX or chiral N2LO 
three-nucleon potential in the case of $A=3$. 
The weak current consists of polar- and axial-vector components. The
former are related to the isovector piece of the electromagnetic
current via the conserved-vector-current hypothesis. These and the
axial currents are derived either in a meson-exchange or in a chiral
effective field theory ($\chi$EFT) framework.
There is one parameter (either the $N$-to-$\Delta$ axial coupling
constant in the meson-exchange model, or the strength of a contact term in the
$\chi$EFT model) which is fixed by reproducing the Gamow-Teller
matrix element in tritium $\beta$-decay. The model dependence
relative to the adopted interactions and currents (and cutoff
sensitivity in the $\chi$EFT currents) is weak, resulting
in total rates of $392.0\pm 2.3$ s$^{-1}$ for $A=2$, and 
$1484\pm 13$ s$^{-1}$  for $A=3$, where the spread accounts for 
this model dependence.

\end{abstract}
\pacs{23.40.-s,21.45.-v,27.10.+h}

\maketitle

\section{Introduction}
\label{sec:intro}

There is a significant body of experimental and theoretical work
on muon capture in nuclei (see Refs.~\cite{Mea01,Gor04} for a review).
These processes provide a testing ground for wave functions and,
indirectly, the interactions from which these are obtained, and
for models of the nuclear
weak current. This is particularly important for neutrino reactions
in light nuclei~\cite{Gazetal} and for processes, such
as the astrophysically relevant weak captures on proton 
and $^3$He, whose rates cannot be measured experimentally,
and for which one has to rely exclusively on theory.
Thus,
it becomes crucial to study within the same theoretical framework
related electroweak transitions, whose rates are known
experimentally~\cite{SFII}. Muon captures
are among such reactions. 

In the present work, we focus our attention on muon capture on 
deuteron and $^3$He, i.e., on the reactions
\begin{eqnarray}
&&\mu^-+d\rightarrow n+n+\nu_\mu \ , \label{eq:mud}\\
&&\mu^-+\,^3{\rm He}\rightarrow\,^3{\rm H}+\nu_\mu \ . \label{eq:mu3}
\end{eqnarray}
Muon capture on $^3$He can also occur through the two- ($nd$) and
three-body ($nnp$) breakup channels of $^3{\rm H}$. However, 
the branching ratios of these two processes are 
20\% and 10\%, respectively, and will not be considered in the 
present work. 

These reactions have been studied extensively through the years, 
experimentally and theoretically. 
In reaction~(\ref{eq:mud}), 
the stopped muons are captured from two hyperfine states, $f=1/2$ or $3/2$.
The doublet capture rate $\Gamma^D$ has been calculated by several groups 
to be about 40 times larger than the quadruplet~\cite{Mea01,Gor04}.
We will therefore consider only $\Gamma^D$.
The first attempt to measure $\Gamma^D$ was carried out over
forty years
ago by Wang {\it et al.}: they obtained 
$\Gamma^D=365(96)$ s$^{-1}$~\cite{Wan65}.
A few years later, Bertin {\it et al.} measured
$\Gamma^D=445(60)$ s$^{-1}$~\cite{Ber73}. 
Measurements performed in the eighties gave
$\Gamma^D=470(29)$ s$^{-1}$~\cite{Bar86}
and $\Gamma^D=409(40)$ s$^{-1}$~\cite{Car86}.
These measurements, while consistent with each other, are not 
very precise -- errors are in the 6--10 \% range. 
However, there
is hope to have this situation clarified by the MuSun 
Collaboration~\cite{And07}, which is expected to 
perform an experiment at the
Paul Scherrer Institut, with the goal of 
measuring $\Gamma^D$ with a precision of 1 \%.

The experimental situation for reaction~(\ref{eq:mu3}) is much clearer: 
after a first set of measurements in the early 
sixties~\cite{Fal63,Zai63,Aue65,Cla65}, 
a very precise determination in the late nineties 
yielded a total capture rate $\Gamma_0$=1496(4) s$^{-1}$~\cite{Ack98},
a value 
consistent with those of the earlier measurements, 
although these were affected by considerably larger uncertainties.

Theoretical work on reactions~(\ref{eq:mud}) and~(\ref{eq:mu3})
is just as extensive, and
a list of publications, updated to the late nineties, is given in
Table 4.1 of Ref.~\cite{Mea01}, and in Ref.~\cite{Gor04}.
Here, we limit our considerations to the calculations 
of Refs.~\cite{Ada90,Tat90,Doi90-91}. 
We also comment 
on the recent studies of Ando {\it et al.}~\cite{And02} and 
Ricci {\it et al.}~\cite{Ric10}. 

The calculations of Refs.~\cite{Ada90,Tat90,Doi90-91} were performed
within the ``Standard Nuclear Physics Approach'' (SNPA):
their authors used the realistic 
potential models available at the time 
to obtain the nuclear
wave functions, and included in the nuclear weak current operator
one-body (impulse approximation) and two-body operators. 
In Ref.~\cite{Ada90}, $\Gamma^D$ was
calculated to be 416(7) s$^{-1}$, the uncertainty coming from imprecise
knowledge of the coupling constants, and in Refs.~\cite{Tat90} 
and~\cite{Doi90-91} 399 s$^{-1}$ and 402 s$^{-1}$, 
respectively. The results of Refs.~\cite{Tat90} and~\cite{Doi90-91}
are in good agreement with each other, while that of Ref.~\cite{Ada90}
differs by $\sim 4$ \%.
It is important to note, however, that 
the meson-exchange currents (MEC) contributions were not constrained to 
reproduce any experimental observable, such as the 
triton half-life, as is now
common practice~\cite{Sch98,Mar00,Par03}.

Reference~\cite{And02} overcame some of the limitations inherent
to the earlier studies. Much along the lines of Ref.~\cite{Par03}, 
reaction~(\ref{eq:mud}) was studied within a hybrid chiral 
effective field theory ($\chi$EFT) approach, 
in which matrix elements of weak operators derived in $\chi$EFT
were evaluated between wave functions obtained from a realistic potential,
specifically the Argonne $v_{18}$ (AV18)~\cite{Wir95}.
The $\chi$EFT axial current contains a low-energy constant
which was fixed by reproducing the experimental Gamow-Teller 
matrix element (GT$^{\rm EXP}$) in tritium $\beta$-decay. 
The calculation, however, 
retained only the $S$-wave contribution in the $nn$
final scattering state (the $^1S_0$ state), and higher partial-wave 
contributions were taken from Ref.~\cite{Tat90}. This approach
yielded a value for $\Gamma^D$ of 386 s$^{-1}$, 
with $\Gamma^D(^1S_0)$=245(1) s$^{-1}$, 
the theoretical error being related to the experimental uncertainty in 
GT$^{\rm EXP}$.

The latest SNPA calculation of muon capture on deuteron has been 
carried out 
in Ref.~\cite{Ric10}, and has led to values
in the range of 416--430 s$^{-1}$ (see Table 1 of Ref.~\cite{Ric10}), 
depending on the potential used, 
the Nijmegen I or Nijmegen 93~\cite{Sto94}. 
However, the model for the axial current is not constrained by data,
resulting in the relatively large spread in $\Gamma^D$ values.

Finally, there is a calculation based on 
pionless EFT~\cite{Che05} with the objective of constraining
the two-nucleon axial current matrix element by reproducing the
muon capture rate on deuterons. This same matrix element
enters the $pp$ weak capture.

Theoretical studies for reaction~(\ref{eq:mu3}) within the 
SNPA have been performed in the early nineties by 
Congleton and Fearing~\cite{Con92} and Congleton and Truhlik~\cite{Con96}.
In this later work, 
the nuclear wave functions were obtained from a realistic 
Hamiltonian based on the Argonne $v_{14}$ (AV14) 
two-nucleon~\cite{Wir84} and the 
Tucson-Melbourne (TM) three-nucleon~\cite{Coo79} interactions. The nuclear 
weak current retained contributions similar to those of Ref.~\cite{Ric10}. 
The value obtained for the total capture rate $\Gamma_0$ 
was 1502(32) s$^{-1}$, the uncertainty due to poor knowledge 
of some of the coupling constants and cutoff parameters entering the 
axial current. 

A first study of the model-dependence of predictions for
the total rate of muon capture on 
$^3$He was carried out in Ref.~\cite{Ho02}, within the SNPA, 
but without the inclusion of MEC contributions. Values for $\Gamma_0$ were
found to vary by $\simeq 100$ s$^{-1}$, depending on the potential 
model considered. Both old- (Paris~\cite{Paris}, Bonn A and B~\cite{BonnB}, 
AV14)
and new-generation (Nijmegen I~\cite{Sto94} and CD-Bonn~\cite{CDBonn})
potentials were used. However,
no three-nucleon forces were included. This second aspect of the calculation, 
along with the absence of MEC contributions,
might be the origin of the large model dependence
observed in the results.

A first attempt to study muon capture on $^3$He 
in a way that was consistent with 
the approach adopted for the weak proton capture reactions, involved
some of the present authors~\cite{Mar02}.
The nuclear wave functions were obtained 
with the hyperspherical-harmonics (HH) method 
(see Ref.~\cite{Kie08} for a recent review),  
from a realistic Hamiltonian based on the AV18 two-nucleon 
and Urbana IX~\cite{Pud95} (UIX) three-nucleon interactions. The model 
for the nuclear weak current was taken from Refs.~\cite{Sch98,Mar00}. 
However, two additional contributions
were included: the single-nucleon pseudoscalar charge 
operator and the pseudoscalar two-body term in the $N$-to-$\Delta$
transition 
axial current. Both contributions are of order $O(q^2/m^2)$, where 
$q$ is the momentum transfer in the process and $m$ is the nucleon mass, and
were obviously neglected in the $pp$ and $hep$ captures
of Refs.~\cite{Sch98,Mar00}, for which $q<<m$. The 
axial coupling constant for the $N$-to-$\Delta$ transition
was constrained to reproduce
GT$^{\rm EXP}$. The total capture rate $\Gamma_0$ 
was found to be 1484(8) s$^{-1}$, where the uncertainty 
results from the adopted 
fitting procedure and experimental error on GT$^{\rm EXP}$. 
A calculation based on 
the older AV14/TM Hamiltonian model yielded a $\Gamma_0$ of 
1486(8) s$^{-1}$,  suggesting 
a weak model-dependence.

Recently, a hybrid calculation has appeared~\cite{Gaz08},
in which
the nuclear wave functions have been obtained with the
Effective Interaction HH method~\cite{EIHH},
and the $\chi$EFT weak current is that of Ref.~\cite{Par03}. 
It has yielded a value for $\Gamma_0$ of 1499(16) s$^{-1}$, 
where the error has two 
main sources: the experimental uncertainty on the triton half-life, and
the calculation of radiative corrections.

In light of the previous considerations, it is clear that a calculation
is still lacking which: (i) treats reactions~(\ref{eq:mud})
and~(\ref{eq:mu3}) simultaneously in a consistent framework,
either SNPA or $\chi$EFT; (ii) is based on
up-to-date Hamiltonian models to generate the wave functions;
(iii) reduces the model dependence of the weak axial current 
by constraining it to reproduce GT$^{\rm EXP}$.
The goal of the 
present work is to fill this gap. The calculation has been structured 
as follows: the nuclear Hamiltonian models considered consist of the
AV18 and N3LO~\cite{Ent03} two-nucleon interactions 
for the $A=2$ systems, augmented by the 
UIX and N2LO~\cite{Nav07} three-nucleon interactions for the 
$A=3$ systems. 
The nuclear weak current is derived from either
SNPA or $\chi$EFT. In both cases, its axial component is
calibrated by fitting GT$^{\rm EXP}$, while its vector part
is related to the isovector electromagnetic current by the
conserved-vector-current (CVC) hypothesis. The SNPA version of it
reproduces well static magnetic properties of few-nucleons
systems. In the $\chi$EFT electromagnetic current
of Ref.~\cite{Son09}, adopted in the present work, the
two low-energy constants are fixed by reproducing the trinucleon
magnetic moments.

In closing, we note that the study of Ref.~\cite{Mar02}
has established that the total rate of reaction~(\ref{eq:mu3})
scales approximately linearly with 
the trinucleon binding energy. 
Thus a realistic calculation of this rate must,
at a minimum, include a three-nucleon potential
which reproduces well these energies (as is the case
here).  Finally, we could have adopted the set of
two-nucleon chiral potentials at N3LO derived by the Bonn 
group~\cite{Epe05} for a range of cutoff parameters $\Lambda$, in order to
explore the sensitivity of the results to the short-range
behavior of the potentials.  However, the potentials considered
in the present study (AV18 and N3LO) have so drastically different treatments
of this short-range behavior that they should provide a meaningful measure
of the model dependence originating from it.

The paper is organized as follows. In Sec.~\ref{sec:obs} we list the explicit
expressions for the observables of interest in terms of reduced matrix elements
of multipole operators. In the case of reaction~(\ref{eq:mu3}) the
derivation is given in Ref.~\cite{Mar02}. In Sec.~\ref{sec:nucwf}, we
briefly review the method used 
to calculate the nuclear wave functions and summarize the 
main results for the observables of the nuclear systems involved 
in the reactions of interest. In Sec.~\ref{sec:weak}, we 
describe the model for the nuclear weak current, both its SNPA 
and $\chi$EFT versions. In Sec.~\ref{sec:res}, we present and discuss  
the results obtained for the total capture rates 
of reactions~(\ref{eq:mud}) and~(\ref{eq:mu3}), and finally
in Sec.~\ref{sec:sum} we summarize our conclusions.

\section{Observables}
\label{sec:obs}

The muon capture on deuteron and $^3$He is induced by the weak interaction
Hamiltonian~\cite{Wal95}
\begin{equation}
H_{W}={G_{V}\over{\sqrt{2}}} \int {\rm d}{\bf x}
\, l_{\sigma}({\bf x}) j^{\sigma}({\bf x}) \ , \label{eq:hw}
\end{equation}
where $G_{V}$ is the Fermi coupling constant,
$G_{V}$=1.14939 $\times 10^{-5}$ GeV$^{-2}$ as
obtained from an analysis of $0^+$ to $0^+$ $\beta$-decays~\cite{Har90},
and $l_\sigma$ and $j^\sigma$ are the leptonic and
hadronic current densities, respectively.  The former
is given by
\begin{equation}
l_{\sigma}({\bf x}) =\, 
{\rm e}^{-{\rm i} {\bf k}_\nu \cdot {\bf x} } \,
{\overline{u}}({\bf k}_\nu,h_\nu)\,\gamma_{\sigma}\, (1-\gamma_5)
\psi_{\mu}({\bf x},s_{\mu}) \>\>\>,
\label{eq:lepc}
\end{equation}
where $\psi_\mu({\bf x},s_\mu)$ is the ground-state
wave function of the muon in the Coulomb field of the nucleus in the
initial state, and $u({\bf k}_\nu,h_\nu)$ is the spinor of
a muon neutrino with momentum ${\bf k}_\nu$, energy
$E_\nu$ (=$k_\nu$), and helicity $h_\nu$.  While in principle
the relativistic solution of the Dirac equation could be
used, in practice it suffices to approximate
\begin{eqnarray}
\psi_\mu({\bf x},s_\mu) &\simeq& 
\psi_{1s}(x) \chi(s_\mu) \equiv 
\psi_{1s}(x) u({\bf k}_\mu,s_\mu) \nonumber \\
{\bf k}_\mu &\rightarrow& 0 \>\>\>,
\label{eq:psimu}
\end{eqnarray}
since the muon velocity
$v_\mu \simeq Z \alpha \ll 1$ ($\alpha$ is
the fine-structure constant and $Z$=1 or 2 for deuteron or
$^3$He, respectively).  Here
$\psi_{1s}(x)$ is the $1s$ solution of the Schr\"odinger
equation and, since the muon is essentially at rest,
it is justified to replace the two-component
spin state $\chi(s_\mu)$ with the four-component
spinor $u({\bf k}_\mu,s_\mu)$
in the limit ${\bf k}_\mu \rightarrow 0$.  This will
allow us to use standard techniques to carry out
the spin sum over $s_\mu$ at a later stage. 

In order to account for the hyperfine structure
in the initial system, the muon and deuteron or $^3$He
spins are coupled to states with total spin $f$, equal to 1/2 or 3/2 in the
deuteron case, and to 0 or 1 in the $^3$He case.  
The transition amplitude can then be conveniently
written as
\begin{widetext}
\begin{eqnarray}
T_W (f,f_z;s_1,s_2,h_\nu) &\equiv&
\langle nn, s_1, s_2; \nu, h_\nu \,|\, H_W \,|\,
(\mu,d);f,f_z \rangle
\nonumber \\
&\simeq& {G_V \over \sqrt{2}} \psi_{1s}^{\rm av}
\sum_{s_\mu s_d}
\langle {1 \over 2}s_{\mu}, 1 s_d | f f_z \rangle\,
l_\sigma(h_\nu,\,s_\mu)\,
\langle \Psi_{{\bf p}, s_1 s_2}(nn) | j^{\sigma}({\bf q}) |
\Psi_d(s_d)\rangle \ , \label{eq:h2ffz}
\end{eqnarray}
\end{widetext}
for the muon capture on deuteron, where ${\bf p}$ is the $nn$ relative
momentum, and~\cite{Mar02}
\begin{widetext}
\begin{eqnarray}
T_W (f,f_z;s^\prime_{3},h_\nu) &\equiv&
\langle ^3{\rm H}, s^\prime_{3}; \nu, h_\nu \,|\, H_W \,|\,
(\mu,^3\!{\rm He});f,f_z \rangle
\nonumber \\
&\simeq& {G_V \over \sqrt{2}} \psi_{1s}^{\rm av}
\sum_{s_\mu s_3}
\langle {1 \over 2}s_{\mu}, {1 \over 2} s_3 | f f_z \rangle\,
l_\sigma(h_\nu,\,s_\mu)\,
\langle \Psi_{^3{\rm H}}(s^\prime_{3}) | j^{\sigma}({\bf q}) |
\Psi_{^3{\rm He}} (s_3)\rangle \ , \label{eq:h3ffz}
\end{eqnarray}
\end{widetext}
%
for muon capture on $^3$He.
In Eqs.~(\ref{eq:h2ffz}) and~(\ref{eq:h3ffz}) we have defined 
\begin{equation}
l_\sigma(h_\nu,\,s_\mu) \equiv
{\overline{u}}({\bf k}_\nu,h_\nu)\,\gamma_{\sigma}\, (1-\gamma_5)
u({\bf k}_\mu,s_{\mu}) \>\>\>,
\label{eq:lsigma}
\end{equation}
and the Fourier transform of the nuclear weak current
has been introduced as
\begin{equation}
j^\sigma({\bf q})=\int {\rm d}{\bf x}\,
{\rm e}^{ {\rm i}{\bf q} \cdot {\bf x} }\,j^\sigma({\bf x})
\equiv (\rho({\bf q}),{\bf j}({\bf q}))
\label{eq:jvq} \>\>\>,
\end{equation}
with the leptonic momentum transfer ${\bf q}$ defined
as ${\bf q} = {\bf k}_\mu-{\bf k}_\nu \simeq -{\bf k}_\nu$.
The function $\psi_{1s}(x)$ has been
factored out from the matrix element of $j^{\sigma}({\bf q})$
between the initial and final states. For muon capture on deuteron,
$\psi_{1s}^{\rm av}$ is approximated as~\cite{Wal95}
\begin{equation}
|\psi_{1s}^{\rm av}|^2 \equiv\,  |\psi_{1s}(0)|^2\,=\,
{(\alpha\, \mu_{\mu d})^3\over \pi} \ ,
\label{eq:psimud}
\end{equation}
where $\psi_{1s}(0)$ denotes the Bohr wave function
for a point charge $e$ evaluated at the origin, and
$\mu_{\mu d}$ is the reduced mass of the $(\mu,d)$ system.
For muon capture on $^3$He,
$\psi_{1s}^{\rm av}$ is approximated as~\cite{Mar02}
\begin{equation}
|\psi_{1s}^{\rm av}|^2 \equiv\,  
{\cal {R}}\,{(2\,\alpha\, \mu_{\mu ^3{\rm He}})^3\over \pi} \ ,
\label{eq:psimu3}
\end{equation}
where in this case 
$\mu_{\mu ^3{\rm He}}$ is the reduced mass of the ($\mu,^3$He) system,
and the factor ${\cal {R}}$ approximately accounts
for the finite extent of the nuclear charge
distribution~\cite{Wal95}. 
This factor is defined as
\begin{equation}
{\cal R}=\frac{|\psi_{1s}^{\rm av}|^2}{|\psi_{1s}(0)|^2}\ ,
\label{eq:rf}
\end{equation}
with
\begin{equation}
\psi_{1s}^{\rm av}=\frac
{\int d{\bf x} \,{\rm e}^{{\rm i}{\bf q}\cdot{\bf x}}\psi_{1s}(x)\rho(x)}
{\int d{\bf x} \,{\rm e}^{{\rm i}{\bf q}\cdot{\bf x}}\rho(x)} \ ,
\label{eq:psiav}
\end{equation}
where $\rho(x)$ is the $^3$He charge density. 
It has been calculated
explicitly by using the charge densities corresponding to the
two Hamiltonian models considered in the present study (AV18/UIX and
N3LO/N2LO), and has been found to be, for both models, within a percent
of 0.98, the value commonly adopted in the literature~\cite{Wal95}.

In the case of muon capture on deuteron, the final state wave
function is expanded in partial waves as
\begin{widetext}
\begin{equation}
\Psi_{{\bf p},s_1,s_2}(nn)=4\pi\sum_{S} \langle \frac{1}{2} s_1,\frac{1}{2}s_2 |
S S_z \rangle 
\sum_{L J J_z}{\rm i}^L Y^*_{LL_z}({\hat{\bf p}}) 
\langle S S_z, L L_Z | J J_z\rangle \,\overline{\Psi}_{nn}^{LSJJ_z}(p) \>\> ,
\label{eq:psinnpw}
\end{equation}
\end{widetext}
where $\overline{\Psi}_{nn}^{LSJJ_z}(p)$ is the $nn$ wave function
-- it will be discussed in Sec.~\ref{sec:nucwf}.
In the present work, we restrict our calculation to $J\leq 2$ and
$L\leq 3$, and therefore the contributing partial waves are,
in a spectroscopic notation, $^1S_0$, $^3P_0$, $^3P_1$, $^3P_2$--$^3F_2$ 
and $^1D_2$. 

Standard techniques~\cite{Mar00,Wal95} are
now used to carry out the multipole expansion
of the weak charge, $\rho({\bf q})$, and current,
${\bf j}({\bf q})$, operators. For muon capture on 
deuteron, we find
\begin{widetext}
\begin{eqnarray}
\langle \overline{\Psi}_{nn}^{LSJJ_z}(p) | \rho({\bf q}) | \Psi_d(s_d) \rangle 
&=&
\sqrt{4\pi}\sum_{\Lambda \geq 0}\sqrt{2\Lambda+1}\,\,{\rm i}^\Lambda
\frac{\langle 1 s_d, \Lambda 0 | J J_z\rangle}{\sqrt{2J+1}} C_\Lambda^{LSJ}(q) \ ,
\label{eq:c2} \\
\langle \overline{\Psi}_{nn}^{LSJJ_z}(p) | j_z({\bf q}) | \Psi_d(s_d) \rangle 
&=&
-\sqrt{4\pi}\sum_{\Lambda \geq 0}\sqrt{2\Lambda+1}\,\,{\rm i}^\Lambda
\frac{\langle 1 s_d, \Lambda 0 | J J_z\rangle}{\sqrt{2J+1}} L_\Lambda^{LSJ}(q) \ ,
\label{eq:l2} \\
\langle \overline{\Psi}_{nn}^{LSJJ_z}(p) | j_\lambda({\bf q}) | 
\Psi_d(s_d) \rangle 
&=&
\sqrt{2\pi}\sum_{\Lambda \geq 1}\sqrt{2\Lambda+1}\,\,{\rm i}^\Lambda
\frac{\langle 1 s_d, \Lambda -\lambda | J J_z\rangle}{\sqrt{2J+1}} 
[-\lambda M_\lambda^{LSJ}(q)+E_\Lambda^{LSJ}(q)]\ ,
\label{eq:em}
\end{eqnarray}
\end{widetext}
where $\lambda=\pm 1$, and $C_{\Lambda}^{LSJ}(q)$, $L_{\Lambda}^{LSJ}(q)$,
$E_{\Lambda}^{LSJ}(q)$ and $M_{\Lambda}^{LSJ}(q)$ denote the reduced matrix 
elements (RMEs) of the Coulomb ($C$), longitudinal ($L$), transverse
electric ($E$) and transverse magnetic ($M$) multipole operators, as defined
in Ref.~\cite{Mar00}. Since the weak charge and current
operators have scalar/polar-vector $(V)$ and pseudo-scalar/axial-vector $(A)$
components, each multipole consists of the sum of $V$ and $A$ terms,
having opposite parity under space inversions~\cite{Mar00}.
The contributing multipoles for the $S$-, $P$-, and $D$-channels
considered in the present work are listed in Table~\ref{tab:rme2},
where the superscripts $LSJ$ have been dropped.

In the case of muon capture on $^3$He, explicit expressions for the
multipole operators can be found in Ref.~\cite{Mar02}. 
Here we only note that parity
and angular momentum selection rules restrict the contributing RMEs to
$C_0(V)$, $C_1(A)$, $L_0(V)$, $L_1(A)$, $E_1(A)$, and $M_1(V)$.

The total capture rate for the two reactions under consideration
is then defined as
\begin{equation}
d\Gamma = 2\pi\delta(\Delta E) \overline{|T_W|^2} \times({\rm phase \, space})
\ ,
\label{eq:dgamma}
\end{equation}
where $\delta(\Delta E)$ is the energy-conserving $\delta$-function, 
and the phase space is 
$d{\bf p}\,d{\bf k}_\nu/(2\pi)^6$ 
for reaction~(\ref{eq:mud}) and just
$d{\bf k}_\nu/(2\pi)^3$ for reaction~(\ref{eq:mu3}).
The following notation has been introduced: (i) for muon capture
on deuteron
\begin{equation}
\overline{|T_W|^2} = \frac{1}{2f+1}\sum_{s_1 s_2 h_\nu}\sum_{f_z}
|T_W(f,f_z;s_1,s_2, h_\nu)|^2 \ ,
\label{eq:hw2}
\end{equation}
and the initial hyperfine state has been fixed to be $f=1/2$; 
(ii) for muon capture on $^3$He
\begin{equation}
\overline{|T_W|^2} = \frac{1}{4}\, \sum_{s_3^\prime  h_\nu}\sum_{f  f_z}
|T_W(f,f_z;s^\prime_3, h_\nu)|^2 \ ,
\label{eq:hw3}
\end{equation}
and the factor 1/4 follows from assigning the same probability 
to all different hyperfine states. 

After carrying out the spin sums, 
the total rate for muon capture on $^3$He reads~\cite{Mar02}
\begin{widetext}
\begin{equation}
\Gamma_0 = G_V^2\, E_\nu^2\, \left(1-{E_\nu \over m_{^3{\rm H}}}\right)
\,|\psi_{1s}^{\rm av}|^2 
\Big[\,|C_0(V)-L_0(V)|^2\,+\,|C_1(A)-L_1(A)|^2 
+|M_1(V)-E_1(A)|^2 \,\Big] \ ,
\label{eq:g0}
\end{equation}
\end{widetext}
with $E_\nu$ given by
\begin{equation}
E_\nu=\frac{(m_\mu + m_{^3{\rm He}})^2 -m_{^3{\rm H}}^2}{2 (m_\mu + m_{^3{\rm He}})}
 \ .
\label{eq:enu3}
\end{equation}
In the case of muon capture on deuteron, the differential rate reads
\begin{equation}
\frac{d\Gamma_D}{dp}=E_\nu^2\, \left[ 1-{E_\nu \over (m_\mu + m_d)}\right]
\,|\psi_{1s}^{\rm av}|^2 \frac{p^2d{\hat{\bf p}}}{8\pi^4}\,
\overline{|T_W|^2} \ ,
\label{eq:dgd}
\end{equation}
where
\begin{equation}
E_\nu=\frac{(m_\mu +m_d)^2 -4m_n^2-4 p^2}{2 (m_\mu +m_d)} \ .
\label{eq:enu2}
\end{equation}
In Eqs.~(\ref{eq:g0})--(\ref{eq:enu2}), $m_\mu$, $m_n$, $m_d$, $m_{^3{\rm H}}$,
$m_{^3{\rm He}}$ are the muon, neutron, deuteron, $^3$H and $^3$He masses.
The integration over ${\hat{\bf p}}$ in Eq.~(\ref{eq:dgd})
is performed numerically using Gauss-Legendre points,
and a limited number of them, of the order of 10, 
is necessary to achieve convergence to better than 1 part in 10$^3$. In 
order to
calculate the total capture rate $\Gamma^D$, the differential 
capture rate
is plotted versus $p$, and numerically
integrated. Usually, about 30 points in $p$ are enough for this integration
in each partial wave. The differential cross section for each 
contributing partial wave will be shown in Sec.~\ref{sec:res}.

\begin{center}
\begin{table}[hb]
\begin{tabular}{cc}
\hline
Partial wave & Contributing multipoles \\
\hline
$^1S_0$ & $C_1(A)$, $L_1(A)$, $E_1(A)$, $M_1(V)$ \\
$^3P_0$ & $C_1(V)$, $L_1(V)$, $E_1(V)$, $M_1(A)$ \\
$^3P_1$ & $C_0(A)$, $L_0(A)$, \\
        & $C_1(V)$,  $L_1(V)$,  $E_1(V)$,  $M_1(A)$,\\
        & $C_2(A)$,  $L_2(A)$,  $E_2(A)$,  $M_2(V)$ \\
$^3P_2$--$^3F_2$
        & $C_1(V)$,  $L_1(V)$,  $E_1(V)$,  $M_1(A)$,\\
        & $C_2(A)$,  $L_2(A)$,  $E_2(A)$,  $M_2(V)$,\\
        & $C_3(V)$,  $L_3(V)$,  $E_3(V)$,  $M_3(A)$ \\
$^1D_2$ 
        & $C_1(A)$,  $L_1(A)$,  $E_1(A)$,  $M_1(V)$,\\
        & $C_2(V)$,  $L_2(V)$,  $E_2(V)$,  $M_2(A)$,\\
        & $C_3(A)$,  $L_3(A)$,  $E_3(A)$,  $M_3(V)$ \\ 
\hline
\end{tabular}
\caption{\label{tab:rme2} Contributing
multipoles in muon capture on deuteron, for all the $nn$ partial waves
with $J\leq 2$ and $L\leq 3$.
The spectroscopic notation is used. See text for further explanations.}
\end{table}
\end{center}
\section{Nuclear wave functions}
\label{sec:nucwf}

Bound and continuum wave functions for both two- and three-nucleon
systems are obtained with the hyperspherical-harmonics (HH)
expansion method. This method, as implemented in the case of
$A=3$ systems, has been reviewed in considerable detail in a series
of recent publications~\cite{Kie08,Viv06,Mar09}. We will discuss it here 
in the context of $A=2$ systems, for which, of course,
wave functions could have been obtained by direct solution
of the Schr\"odinger equation. 

The nuclear wave function for a bound system of
total angular momentum $J J_z$ can be generally written as
\begin{equation}
|\Psi^{JJ_z}\rangle=\sum_\mu c_\mu |\Psi^{JJ_z}_\mu\rangle \ ,
\label{eq:psi}
\end{equation}
where $|\Psi^{JJ_z}_\mu\rangle$ is a complete set of 
states, and $\mu$ is an index denoting the set of quantum numbers 
necessary to completely specify the basis elements. 
The coefficients of the expansion can be calculated by using the 
Rayleigh-Ritz variational principle, which states that
\begin{equation}
  \langle\delta_c \Psi^{JJ_z}\,|\,H-E\,|\Psi^{JJ_z}\rangle
   =0 \ ,
   \label{eq:rrvar}
\end{equation}
where $\delta_c \Psi^{JJ_z}$ indicates the variation of 
$\Psi^{JJ_z}$ for arbitrary infinitesimal 
changes of the linear coefficients $c_\mu$. 
The problem of determining $c_\mu$ and the energy $E$ 
is then reduced to a generalized eigenvalue problem, 
\begin{equation}
  \sum_{ \mu'}\,\langle\Psi^{JJ_z}_\mu\,|\,H-E\,|\, 
\Psi^{JJ_z}_{\mu'}\,\rangle \,c_{\mu'}=0
  \ .
  \label{eq:gepb}
\end{equation}
In the case of the deuteron, we first define
\begin{equation}
\psi_{LSJJ_z}({\bf r})=F_{LSJ}(r){\cal Y}_{LSJJ_z}({\hat{\bf r}})\ , 
\label{eq:psi2}
\end{equation}
where
\begin{equation}
{\cal Y}_{ LSJJ_z }({\hat{\bf r}})=[Y_{L}({\hat{\bf r}})\otimes \chi_{S}]_{JJ_z}
\ , \label{eq:y2}
\end{equation}
and ${\bf r}$ is the relative position vector of the two nucleons. 
Obviously, for the deuteron, $L=0$ or 2, $S=1$, $J=1$, and $T=0$ 
(the isospin state $\eta_{TT_z}$ with $T=T_z=0$ has been 
dropped for brevity). The radial functions $F_{LSJ}(r)$ are
then conveniently expanded on a basis of Laguerre polynomials as
\begin{equation}
F_{LSJ}(r)=\sum_l c_{LSJ,l}\,f_l(r)\, \ ,
\label{eq:plag}
\end{equation}
with
\begin{equation}
 f_l(r)=\sqrt{\frac{l!}{(l+2)!}}\,\gamma^{3/2} \,\, 
 L^{(2)}_l(\gamma r)\,\,{\rm e}^{-\gamma r/2} \ ,
 \label{eq:fllag2}
\end{equation}
where the parameter $\gamma$ is 
variationally optimized ($\gamma$ is in the range of 3--4 fm$^{-1}$ for the
AV18 and N3LO potentials).
The complete wave function is then reconstructed as
\begin{eqnarray}
\Psi^{1 J_z}&=&
\sum_{L=0,2}\sum_l c_{L11,l} f_l(r){\cal Y}_{L11J_z}({\hat{\bf r}})
\nonumber \\
&\equiv&\sum_\mu c_\mu \Psi_\mu^{1J_z} \ ,
\label{eq:psi2tot}
\end{eqnarray}
and the subscript $\mu$ denotes the set of quantum numbers $[L,S=1,J=1,l]$.

The $nn$ continuum wave function is written as
\begin{equation}
    \Psi^{LSJJ_z}=\Psi_C^{LSJJ_z}+\Psi_A^{LSJJ_z} \ ,
    \label{eq:psica}
\end{equation}
where $\Psi_C^{LSJJ_z}$ describes the system in the region where the 
two neutrons
are close to each other and their mutual interactions are strong, 
while $\Psi_A^{LSJJ_z}$ describes their relative motion 
in the asymptotic region. The function $\Psi_C^{LSJJ_z}$, which
vanishes in the limit of large
separations, is expanded on Laguerre polynomials as before for the case
of the deuteron, see Eq.~(\ref{eq:psi}), while 
the function $\Psi_A^{LSJJ_z}$ is the appropriate 
asymptotic solution in channel $LSJ$,
\begin{equation}
  \Psi_A^{LSJJ_z}= \sum_{L^\prime S^\prime}
 \bigg[\delta_{L L^\prime} \delta_{S S^\prime} 
\Omega_{L^\prime S^\prime JJ_z}^R
  + {\cal R}^J_{LS,L^\prime S^\prime}(p)
     \Omega_{L^\prime S^\prime JJ_z}^I \bigg] \ ,
  \label{eq:psia}
\end{equation}
where 
\begin{equation}
  \Omega_{LSJJ_z}^{R/I}= R^{R/I}_L(pr){\cal Y}_{LSJJ_z}({\hat{\bf r}})\ ,
  \label{eq:psiom}
\end{equation}
with 
\begin{eqnarray}
R^R_L(pr)&\equiv&\frac{1}{p^L}\,j_L(pr) \ , 
\nonumber \\
R^I_L(pr)&\equiv&p^{L+1}\,f_R(r)\,n_L(pr)
\ ,
\label{eq:risol}
\end{eqnarray}
$p$ being the magnitude of the relative momentum. The functions
$j_L(pr)$ and $n_L(pr)$ are the regular and irregular spherical Bessel
functions, and $f_R(r)=[1-\exp(-b r)]^{2L+1}$ 
has been introduced to regularize $n_L(pr)$ at small values of $r$. 
The trial parameter $b$ is taken as $b=0.25$ fm$^{-1}$. 

The matrix elements ${\cal R}^J_{LS,L^\prime S^\prime}(p)$ and 
the coefficients $c_\mu$ entering the expansion of $\Psi^{LSJJ_z}_C$ 
are determined applying the Kohn variational principle~\cite{kohn}, 
stating that the functional
\begin{eqnarray}
&&   [{\cal R}^J_{LS,L^\prime S^\prime}(p)]=
    {\cal R}^J_{LS,L^\prime S^\prime}(p) \nonumber \\
&&\,\,\,\,\,\,\,\,- 
\frac{2\mu}{\hbar^2}\,\left \langle \Psi^{L^\prime S^\prime JJ_z } \left |
         H- \frac{p^2}{2\mu}\right |
        \Psi^{LSJJ_z}\right \rangle \ , \label{eq:kohn}
\end{eqnarray}
is stationary with respect to variations of the trial parameters 
in $\Psi^{LSJJ_z}$. 
Here $\mu$ is the 
reduced mass of the $nn$ system. Performing the variation, a system
of linear inhomogeneous equations for $c_\mu$ and a set of algebraic
equations for ${\cal R}^J_{LS,L^\prime S^\prime}(p)$ are derived, and solved 
by standard techniques. From the ${\cal R}^J_{LS,L^\prime S^\prime}(p)$'s, 
phase shifts, mixing angles, and scattering lengths are easily
obtained.

We list in Tables~\ref{tab:a2} and~\ref{tab:a3} binding energies and 
scattering lengths calculated with the Hamiltonian models considered in the 
present work. Note that $A=3$ wave functions retain both $T=1/2$ and 
$T=3/2$ contributions, $T$ being the total isospin quantum number.
The experimental data are, in 
general, quite well reproduced. It should 
be noted that, by using only two-nucleon interaction, the triton and 
$^3$He binding energies are 7.624 MeV and 6.925 MeV with the AV18, 
and 7.854 MeV and 7.128 MeV with the N3LO, respectively.

\begin{center}
\begin{table}[thb]
\begin{tabular}{cccc}
\hline
   & AV18 & N3LO & Exp. \\
\hline
$B_d$ (MeV) & 2.22457 & 2.22456 & 2.224574(9) \\
$P_D$ (\%)& 5.76 & 4.51 & -- \\
$a_{nn}$ (fm) & --18.487 & --18.900 & --18.9(4) \\
$^1a_{np}$ (fm) & --23.732 & --23.732 & --23.740(20) \\
$^3a_{np}$ (fm) & 5.412 & 5.417 & 5.419(7) \\
\hline
\end{tabular}
\caption{\label{tab:a2}Deuteron binding energy $B_d$ (in MeV) and 
$D$-state probability (in \%), $nn$ and singlet and triplet
$np$ scattering lengths (in fm), calculated with the two-nucleon 
potentials AV18 and N3LO.
The experimental results are from Ref.~\cite{Ent03}.}
\end{table}
\end{center}

Finally, we define
\begin{equation}
\overline{\Psi}^{LSJJ_z}=\sum_{L^\prime S^\prime}
[\delta_{L,L^\prime}\delta_{S,S^\prime}-{\rm i}\,
{\cal R}^J_{LS,L^\prime S^\prime}(p)]^{-1}
\Psi^{L^\prime S^\prime J J_z} \ ,
\label{eq:psiuf}
\end{equation}
so that $\overline{\Psi}^{LSJJ_z}$ has unit flux. The function
$\overline{\Psi}^{LSJJ_z}$ enters in Eq.~(\ref{eq:psinnpw}) and 
in the expression for differential muon capture rate on deuteron.

\begin{center}
\begin{table}[bth]
\begin{tabular}{cccc}
\hline
   & AV18/UIX & N3LO/N2LO & Exp. \\
\hline
$B_{^3{\rm H}}$ (MeV) & 8.479 & 8.474 & 8.482 \\
$B_{^3{\rm He}}$ (MeV) & 7.750 & 7.733 & 7.718 \\
$^2a_{nd}$ (fm) & 0.590 & 0.675 & 0.645(3)(7) \\
$^4a_{nd}$ (fm) & 6.343 & 6.342 & 6.35(2) \\
\hline
\end{tabular}
\caption{\label{tab:a3}Triton and $^3$He binding energies $B_{^3{\rm H}}$ 
and $B_{^3{\rm He}}$ (in MeV), and $nd$ doublet and quartet scattering lengths
(in fm), calculated with the two- and three-nucleon potentials 
AV18/UIX and N3LO/N2LO.
The experimental results are from Ref.~\cite{Kie08}.}
\end{table}
\end{center}

\section{Weak transition operator}
\label{sec:weak}

The nuclear weak charge and current 
operators consist of polar- and axial-vector components.
In the present work, we consider two different models, 
both of which have been used in studies
of weak $pp$ and $hep$ capture reactions in the energy regime relevant to
astrophysics~\cite{Sch98,Mar00,Par03}.
The first model has been 
developed within the so-called ``Standard Nuclear Physics Approach'' (SNPA)
and has been applied also to study weak transitions for $A=6$ and 7 
nuclei~\cite{Sch02}, and  
magnetic moments and $M1$ widths of nuclei with 
$A\leq 7$~\cite{Mar05,Mar08}.
It will be discussed in Sec.~\ref{subsec:snpa}.
In the second model, the nuclear weak transition operators have 
been derived in heavy-baryon chiral perturbation theory (HBChPT),
carrying out the expansion up to next-to-next-to-next-to leading 
order (N3LO)~\cite{Par03,Son07}. This same model has been updated and 
most recently used 
in the electromagnetic sector to study the three-nucleon magnetic moments
and $np$ and $nd$ radiative capture reactions in Ref.~\cite{Son09}.
We review it in Sec.~\ref{subsec:eft}.

\subsection{The ``Standard Nuclear Physics Approach''}
\label{subsec:snpa}

The one-body axial charge and current operators have the standard 
expressions~\cite{Mar00} obtained from the nonrelativistic 
reduction of the covariant single-nucleon current, 
and include terms proportional to $1/m^2$, $m$ being the 
nucleon mass. The induced pseudoscalar contributions
are retained both in the axial current and charge operators, 
and are given by
\begin{eqnarray}
\rho_{i,PS}^{(1)}({\bf q};A)&=& -{g_{PS}(q_\sigma^2)\over{2\,m\,m_\mu}}\,\tau_{i,-}
\,(m_\mu-E_\nu)\nonumber \\
&\times& ({\bm \sigma}_i\cdot{\bf q}) 
\, {\rm e}^{ {\rm i} {\bf q} \cdot {\bf r}_i } \ , \label{eq:rhops}\\
{\bf j}^{(1)}_{i,PS}({\bf q};A)&=&-{g_{PS}(q_\sigma^2) \over {2\, m\, m_\mu}}\, 
\tau_{i,-} 
{\bf q} \, ({\bm \sigma}_i \cdot {\bf q})
\, {\rm e}^{ {\rm i} {\bf q} \cdot {\bf r}_i } \ , \label{eq:jps}
\end{eqnarray}
where 
\begin{equation}
\tau_{i,-}=(\tau_{i,x}-{\rm i}\tau_{i,y})/2 \ , \label{eq:taum}
\end{equation}
and ${\bf q}$ is the three-momentum transfer, 
$m_{\mu}$ the muon mass, $E_\nu$ the neutrino energy, 
$q_\sigma^2$ the squared four-momentum transfer.
Note that, since in the present case ${\bf q}$ is not negligible
($q\simeq m_\mu$), axial and induced 
pseudoscalar form factors need be included. 
In the notation of Ref.~\cite{Mar02}, they are taken as
\begin{eqnarray}
g_A(q_\sigma^2) &=& { g_A \over {(1+ q_\sigma^2/\Lambda_A^2)^2} } \ ,
\label{eq:ga} \\
g_{PS}(q_\sigma^2) &=& -{2\,m_\mu\,m \over m_\pi^2 + q_\sigma^2 }\,
g_{A}(q_\sigma^2) \ .
\label{eq:gps}
\end{eqnarray}
For the axial-vector coupling constant $g_A$, two values have been adopted
in this study:
the first one, $g_A$=1.2654(42), is taken from Ref.~\cite{Ade98}, 
and has been used widely in studies of weak 
processes~\cite{Mar00,Par03,Mar02,Sch02}. The second one,
$g_A$=1.2695(29), is the latest determination quoted by the Particle
Data Group (PDG)~\cite{PDG}. The two values for $g_A$ are consistent 
with each other. 
In order to compare with the results of the aforementioned studies,
we have adopted the earlier determination for $g_A$.
On the other hand, 
to estimate the theoretical uncertainty arising from 
this source,
the calculation has been carried out also using $g_A$=1.2695(29)
in one specific case. As it will be shown below, the central value
and the theoretical uncertainties of the considered observables
are comparable in both cases. 

The value for the cutoff mass $\Lambda_A$ used in this work is 
1 GeV/$c^2$, as in Ref.~\cite{Mar02}. It is obtained
from an analysis of pion electroproduction data~\cite{Ama79}
and measurements of the reaction $\nu_\mu+p\rightarrow n+\mu^+$~\cite{Kit83}.
Since here $q^2_\sigma<<\Lambda_A^2$, an uncertainty of few \% on
$\Lambda_A$ is expected to affect $g_A(q_\sigma^2)$ at the percent level.
The $q^2_\sigma$-dependence of $g_{PS}$ is obtained in accordance with 
the partially-conserved-axial-current (PCAC) hypothesis, by assuming
pion-pole dominance and the Goldberger-Treiman 
relation~\cite{Hem95,Wal95}. In Eq.~(\ref{eq:gps}), 
$m_\pi$ is the pion mass.

The two-body weak axial-charge operator includes a pion-range term,
which follows from soft-pion theorem and current algebra
arguments~\cite{Kub78,Tow92}, and short-range terms,
associated with scalar- and vector-meson exchanges. 
The latter are obtained consistently with the two-nucleon
interaction model, following a procedure~\cite{Kir92} similar
to that used to derive the corresponding weak vector-current
operators~\cite{Mar00}.  The two-body axial charge operator 
due to $N$-to-$\Delta$-transition has also been included~\cite{Mar02,Mar00},
although its contribution is found to be very small.

The two-body axial current operators can be divided in two
classes: the operators of the first class
are derived from $\pi$- and $\rho$-meson exchanges and the 
$\rho\pi$-transition mechanism.  These mesonic operators, 
first obtained in a systematic way
in Ref.~\cite{Che71}, give rather small contributions~\cite{Mar00}.
The operators in the second class are those that give
the dominant two-body contributions, and are due to
$\Delta$-isobar excitation~\cite{Sch98,Mar00}.  We review them 
here briefly.  The $N$-to-$\Delta$-transition axial current is written as
(in the notation of Ref.~\cite{Mar00})
\begin{eqnarray}
{\bf j}_{i}^{(1)}({\bf q}; N\rightarrow\Delta, A) &=& 
- \Bigg[g_A^*(q_\sigma^2) {\bf S}_i +
         \frac{g_{PS}^*(q_\sigma^2)}{2\,m\,m_\mu}{\bf q}
         ({\bf S}_i\cdot{\bf q})\Bigg]
\nonumber \\
         &\times&{\rm e}^{{\rm i}{\bf q}\cdot{\bf r}_i} T_{i,\pm} 
         \ ,
\label{eq:j1ND}
\end{eqnarray}
where ${\bf S}_i$ and ${\bf T}_i$ are spin- and isospin-transition 
operators, which convert a nucleon into a $\Delta$-isobar.  The induced
pseudoscalar contribution has been 
obtained from a non-relativistic reduction of the 
covariant $N$-to-$\Delta$-transition axial current~\cite{Hem95}. 

The axial and pseudoscalar form factors $g_A^*$ and $g_{PS}^*$ 
are parameterized as 
\begin{eqnarray}
g_A^*(q_\sigma^2)&=& R_A\, g_A(q_\sigma^2) \ , \nonumber \\
g_{PS}^*(q_\sigma^2)&=&-{2\,m_\mu\,m \over m_\pi^2 + q_\sigma^2 }\,
g_{A}^*(q_\sigma^2) \ , 
\label{eq:gpsd}
\end{eqnarray}
with $g_A(q_\sigma^2)$ given in Eq.~(\ref{eq:ga}).
The parameter $R_A$ is adjusted to reproduce the experimental 
value of the Gamow-Teller matrix element in tritium $\beta$-decay 
(GT$^{\rm EXP}$), while the $q_\sigma^2$-dependence of $g_{PS}^*$ is again 
obtained by assuming pion-pole dominance and PCAC~\cite{Hem95,Wal95}. 
The value for GT$^{\rm EXP}$, estimated in Ref.~\cite{Sch98}, 
was 0.957(3). This value was determined by assuming $g_A$=1.2654(42),
$\langle {\bf F} \rangle^2$=0.9987, where $\langle {\bf F} \rangle$ 
is the reduced matrix element of the Fermi operator
$\sum_i \tau_{i,-}$, and the triton half-life $fT_{1/2}$ is 
(1134.6$\pm$3.1) s~\cite{Sim87}.
We adopt it also in the present work, except for the 
case in which $g_A$ is taken from the PDG, $g_A$=1.2695(29).
We then extract GT$^{\rm EXP}$ = 0.955(2), 
corresponding to $\langle {\bf F} \rangle^2$=0.99926 and 
$fT_{1/2}$=(1132.1$\pm$4.3) s. The new value of $\langle {\bf F} \rangle^2$
differs by less than 0.1 \% from the older, presumably
due to the higher accuracy of the present trinucleon wave 
functions.
The new value of $fT_{1/2}$ has been obtained by averaging the previous 
value of $fT_{1/2}$ with the new one of Ref.~\cite{Aku05}, (1129.6$\pm$3) s,
and summing the errors in quadrature. 
The values for $R_A$ determined in the present study 
using trinucleon wave functions corresponding to the AV18/UIX 
are $R_A$=1.21(9), when GT$^{\rm EXP}$ = 0.957(3)
and $g_A$=1.2654(42), and $R_A$=1.13(6),
when GT$^{\rm EXP}$ = 0.955(2) and $g_A$=1.2695(29).
The experimental error on GT$^{\rm EXP}$ is responsible for the
5--8 \% uncertainty in $R_A$.

It is important to note that the value of $R_A$ depends
on how the $\Delta$-isobar degrees of freedom are treated.
In the present work, as in Ref.~\cite{Mar02}, the two-body 
$\Delta$-excitation axial operator is
derived in the static $\Delta$ approximation, using first-order
perturbation theory (PT).  This approach
is considerably simpler than that adopted in Ref.~\cite{Mar00},
where the $\Delta$ degrees of freedom were treated non-perturbatively,
within the so-called transition-correlation operator (TCO) approach,
by retaining them explicitly in the nuclear wave functions~\cite{Sch92}.
The results for $R_A$ obtained within the two schemes differ by more 
than a factor of 2~\cite{Mar00}.
However, the results for the observables
calculated consistently within the two different approaches are typically
within 1 \% of each other.
Finally, because of the procedure adopted to determine $R_A$,
the coupling constant $g_A^*$ of Eq.~(\ref{eq:gpsd}) cannot be naively
interpreted as the $N$$\Delta$ axial coupling constant.  The
excitation of additional resonances and their associated
contributions will contaminate the value of $g_A^*$.  

The weak vector charge and current operators are 
constructed from the isovector part of the electromagnetic current,
in accordance with the conserved-vector-current (CVC) hypothesis.
The weak charge operator includes the non-relativistic one-body 
term and the relativistic spin-orbit and Darwin-Foldy contributions, 
and is obtained from the corresponding isovector electromagnetic 
operator, as listed in Ref.~\cite{Car98},
by replacing 
\begin{equation}
\tau_{i,z}/2\rightarrow\tau_{i,-} \ .
\label{eq:cvc1b}
\end{equation}
Two-body contributions,
arising from $\pi$- and $\rho$-meson exchange mechanisms,
are also included. Their expressions are listed in Ref.~\cite{Car98},
with the substitution 
\begin{equation}
({\bm \tau}_i \times {\bm \tau}_j)_z\rightarrow
({\bm \tau}_i \times {\bm \tau}_j)_- \ , 
\label{eq:cvc2b}
\end{equation}
and
\begin{equation}
({\bm \tau}_i \times {\bm \tau}_j)_- = ({\bm \tau}_i \times {\bm \tau}_j)_x
-{\rm i} ({\bm \tau}_i \times {\bm \tau}_j)_y
\ . \label{eq:tauv}
\end{equation}
Electromagnetic form factors are also included, and
available parametrizations for them all provide excellent fits of 
the experimental data at the low momentum transfer of interest here.

The weak vector current operator 
retains the one-body operator, 
two-body ``model-independent'' (MI)
and ``model-dependent'' (MD) terms, and three-body terms. The MI
two-body currents are obtained from the two-nucleon interaction, and by
construction satisfy current conservation with it. In the present work,
we include the leading two-body
``$\pi$-like'' and ``$\rho$-like'' operators, 
obtained from the isospin-dependent
central, spin-spin and tensor nucleon-nucleon interactions. 
On the other hand, we have neglected 
the additional two-body 
currents arising from non-static (momentum-dependent) interactions, since
these
currents are short-ranged, and numerically far less important
than those considered here~\cite{Mar05}.
The MD currents are purely transverse, and therefore cannot be directly
linked to the underlying two-nucleon interaction. 
The present
calculation includes the isovector currents associated with the 
$\omega\pi\gamma$ transition mechanism~\cite{Car98}, 
and the excitation
of $\Delta$ isobars. The former contributions are numerically
negligible, while the latter are important 
in order to reproduce the three-nucleon magnetic moments and
elastic form factors~\cite{Mar05}. 
The contributions of the (MD) $\Delta$-isobar currents 
have been calculated in this case with the TCO method~\cite{Sch92},
and explicit expressions for these are listed in Ref.~\cite{Viv96}. 
Again, the substitution
of Eq.~(\ref{eq:tauv}) is used. 

A three-nucleon interaction requires a corresponding three-body current.
The latter was first derived in Ref.~\cite{Mar05} from a three-nucleon
interaction consisting of a dominant two-pion-exchange component,
such as the UIX model adopted in the present work.
The charge-changing three-body (weak vector) current is obtained 
applying CVC to the operators listed in Ref.~\cite{Mar05},
i.e.
\begin{equation}
\left [ \frac{1}{2}(\tau_{i,a}+\tau_{j,a}+\tau_{k,a})\> ,
 \> {\bf j}^{(3)}_{ijk,z}({\bf q};\gamma) \right ] = {\rm i} \> \epsilon_{azb}\>
{\bf j}^{(3)}_{ijk,b}({\bf q};{\rm V}) \ ,
\end{equation}
where ${\bf j}^{(3)}_{ijk,z}({\bf q};\gamma)$
are the isovector 
three-body electromagnetic currents, and $a,b=x,y,z$
are isospin Cartesian components. 

We conclude by emphasizing that the model for the weak transition
operator is the same as that of Ref.~\cite{Mar02}, 
but for two differences relative to its vector component:
(i) the MD two-body currents due to $\Delta$-isobar degrees of 
freedom are treated non perturbatively with the TCO method,
rather than in first order PT, and (ii) three-body
terms are also included -- they were neglected in Ref.~\cite{Mar02}.
As shown in Table~\ref{tab:mm}, the present model for the electromagnetic
current provides an excellent description of the trinucleon magnetic
moments, in particular of their isovector contribution,
to better than 1 \% (row labeled
with ``FULL''). This gives us confidence in the accuracy
of the corresponding weak vector currents. We also note that:
(i) three-body contributions are important 
to achieve this level of agreement between theory and experiment; 
(ii) the results obtained with the model of Ref.~\cite{Mar02}, 
labeled by ``MD/$\Delta$-PT'', are at variance with 
data at the 3 \% level; (iii) the difference between the present ``FULL''
results and those of Ref.~\cite{Mar05} can be traced back to the 
fact that here we have used slightly more accurate wave functions, 
with a careful treatment of 
$T=3/2$ components. However, we should note that, in contrast 
to Ref.~\cite{Mar05}, MI 
two-body contributions arising from non-static components of
the two-nucleon interaction have been ignored. They amount to
a negligible 0.1 \% correction.
\begin{center}
\begin{table}[tb]
\begin{tabular}{ccc}
\hline
 & $^3$H & $^3$He \\
\hline
IA & 2.5745 & --1.7634 \\
MI & 2.8954 & --2.0790 \\
MD/$\Delta$-PT & 3.0260 & --2.2068 \\
MD/$\Delta$-TCO & 2.9337 & --2.1079 \\
FULL & 2.9525 & --2.1299 \\
Ref.~\protect\cite{Mar05} & 2.953 & --2.125 \\
\hline
Exp. & 2.9790 & --2.1276 \\
\hline
\end{tabular}
\caption{\label{tab:mm} Triton and $^3$He magnetic moments, 
in nuclear magnetons, calculated with the AV18/UIX 
Hamiltonian model and compared with the experimental data. 
The results labeled ``IA'' are obtained with 
single-nucleon current, while those labeled ``MI'' 
are obtained by including in addition 
the model-independent two-body contributions. The results 
labeled ``MD/$\Delta$-PT''
and ``MD/$\Delta$-TCO'' include also 
the model-dependent contributions, with the $\Delta$-isobar degrees of freedom 
treated in perturbation theory or within the TCO approach, respectively. 
Lastly, the results labeled ``FULL'' retain three-body contributions.
Also shown are the results of Ref.~\protect\cite{Mar05}.}
\end{table}
\end{center}

\subsection{The chiral effective field theory approach}
\label{subsec:eft}

The $\chi$EFT weak transition operator 
is taken from Refs.~\cite{Par03} and~\cite{Son09}, where it was
derived in covariant perturbation theory based on the heavy-baryon
formulation of chiral Lagrangians by retaining corrections 
up to N3LO. It was recently used in Ref.~\cite{Laz09} to study
$M1$ electromagnetic transitions. We review its main features here.

The vector and axial-vector one-body operators
are the same as those obtained within the SNPA, and 
described in Sec.~\ref{subsec:snpa}. These one-body operators are 
also listed in Eq.~(17) of Ref.~\cite{Par03},
except that we also include corrections in the vector current,
which arise when the non-relativistic reduction is
carried out to next-to-leading order (proportional to 1/$m^3$).
The resulting operator is given (in momentum space) by
\begin{equation}
{\bf j}_i^{(1)}({\bf q};V)={\bf j}_i^{(1)NR}({\bf q};V)
+ {\bf j}_i^{(1)RC}({\bf q};V)\ ,
\label{eq:jv1b}
\end{equation}
where ${\bf j}_i^{(1)NR}({\bf q};V)$ 
is the standard leading order term
\begin{equation}
{\bf j}_i^{(1)NR}({\bf q};V)=\frac{1}{m}\tau_{i,-}
\Bigl[G_E^V(q_\sigma^2){\bf k}+{\rm i}G_M^V(q_\sigma^2)
{\bm \sigma}_i\times{\bf q}\Bigr] \ ,
\label{eq:jv1bnr}
\end{equation}
and ${\bf j}_i^{(1)RC}({\bf q};V)$ is 
\begin{widetext}
\begin{equation}
{\bf j}_i^{(1)RC}({\bf q};V)= -\frac{1}{4 m^3}\tau_{i,-}
\Bigl[G_E^V(q_\sigma^2)\,{\bf k}^2
(2{\bf k}+{\rm i}{\bm \sigma}_i\times{\bf q}) 
+{\rm i}[G_M^V(q_\sigma^2)-G_E^V(q_\sigma^2)]
(2 {\bf k}\cdot{\bf q}\,{\bm \sigma}_i\times{\bf k}
+ {\bf k}\times{\bf q}\,{\bm \sigma}_i\cdot{\bf k})
\Bigl] \ .
\label{eq:jv1brc}
\end{equation}
\end{widetext}
In the above expressions, 
${\bf k}=({\bf p}'_i+{\bf p}_i)/2$, ${\bf q}={\bf p}'_i-{\bf p}_i$, 
${\bf p}'_i$ and ${\bf p}_i$ being the 
momenta of the outgoing and ingoing nucleons, respectively, and
$G_E^V(q_\sigma^2)$ and $G_M^V(q_\sigma^2)$ denote the isovector combinations 
of the nucleon 
electric and magnetic form factors~\cite{Car98}.
Note that Eqs.~(\ref{eq:jv1b})--(\ref{eq:jv1brc}) 
can also be obtained from Eq.~(11) of 
Ref.~\cite{Son09},
with the substitution~(\ref{eq:cvc1b}) required by CVC.

The axial two-body charge and currents are from Ref.~\cite{Par03}.
In particular, the axial charge
operator is that derived originally in Ref.~\cite{Kub78}.
In the SNPA, additional contributions
are considered, which, in a $\chi$EFT context, are expected to appear 
at higher orders.
The two-body axial current operator consists of two contributions:
a one-pion exchange term and a (non-derivative) two-nucleon 
contact-term. The explicit expressions for these terms 
can be found in Ref.~\cite{Par03}.
While the coupling constants which appear in the one-pion exchange
term
are fixed by $\pi N$ data, the low-energy constant $d_R$, 
determining the strength of the contact-term, has been fixed
by reproducing GT$^{\rm EXP}$. The value $g_A$=1.2654(42) has been used.
The values for $d_R$ are presented in Table~\ref{tab:eft} and discussed 
below.

The two-body vector currents are decomposed into four terms~\cite{Son09}: 
the soft-one-pion-exchange ($1\pi$) term, vertex corrections to the one-pion
exchange ($1\pi C$), the two-pion exchange ($2\pi$), and a contact-term
contribution. Their explicit expressions can be found in Eqs.~(12)--(18)
of Ref.~\cite{Son09}, with the replacements~(\ref{eq:taum}) and~(\ref{eq:tauv})
in the isospin operators. All the
$1\pi$, $1\pi C$ and $2\pi$ contributions contain low-energy constants
estimated in Ref.~\cite{Par95-98-00}, using resonance saturation 
arguments. The contact-term 
electromagnetic contribution is given by
\begin{eqnarray}
{\bf j}^{(2)CT}_{ij}({\bf q};\gamma) &=& -\frac{\rm i}{2m}\,
{\rm e}^{{\rm i}\,{\bf q}\cdot{\bf R}} {\bf q}\times
[\, g_{4S}({\bm \sigma}_i+{\bm \sigma}_j)\nonumber \\
&+& g_{4V} ({\bm \tau}_i\times {\bm \tau}_j)_z 
{\bm \sigma}_i\times {\bm \sigma}_j ] \delta_\Lambda(r) \ .
\label{eq:jvctg}
\end{eqnarray}
The function $\delta_\Lambda(r)$, as well as the Yukawa functions which 
appear in the $1\pi$, $1\pi C$ and $2\pi$ operators and in the two-body 
axial current terms, are obtained by performing the Fourier transform from 
momentum- to coordinate-space with a Gaussian regulator characterized
by a cutoff $\Lambda$. This cutoff determines the momentum scale below which
these EFT currents are expected to be valid, i.e., 
$\Lambda$=500--800 MeV~\cite{Par03}.
The explicit expression of $\delta_\Lambda(r)$ in Eq.~(\ref{eq:jvctg}) is
\begin{equation}
\delta_\Lambda(r)=\int \frac{d{\bf k}^3}{(2\pi)^3}{\rm e}^{-k^2/\Lambda^2}\,
{\rm e}^{{\rm i}{\bf k}\cdot{\bf r}}\ .
\label{eq:delta}
\end{equation}

The coefficients $g_{4S}$ and $g_{4V}$ of Eq.~(\ref{eq:jvctg})
are fixed
to reproduce the experimental values of triton and $^3$He magnetic 
moments, for each nuclear interaction and cutoff value.
This procedure is similar to that used to fix 
the strength $d_R$ of the contact-term in the two-body axial current 
discussed above. The results are presented in Table~\ref{tab:eft}.
A few comments are in order: (i) while values of both
$g_{4S}$ and $g_{4V}$ are presented, only
$g_{4V}$ is relevant in the present work, since CVC relates
the isovector electromagnetic current to the weak vector 
current. (ii) The uncertainty on $g_{4S}$ and $g_{4V}$ is not due 
to the experimental errors on the triton and $^3$He 
magnetic moments, which are in fact negligible, 
but rather to numerics.
We have used a random walk consisting
of 1.6 million configurations, in order to reduce 
the numerical uncertainty on $\Delta\mu$, the difference
between the experimental magnetic moments and the result obtained without
the contact contributions, to less than 1 \%. In contrast,
the experimental error on 
GT$^{\rm EXP}$ is primarily responsible for the
uncertainty in $d_R$.
(iii) The values reported for $d_R$ are different, but consistent
within the error, with those listed in Table II of Ref.~\cite{Par03}.
This is due to differences between the present $^3$H and $^3$He wave 
functions and those of Ref.~\cite{Par03} -- we have already commented
on this in the previous section.
(iv) The $g_{4S}$ and $g_{4V}$ values are 
rather different from those listed in Ref.~\cite{Laz09}. 
We observe that $g_{4S}$ and $g_{4V}$ 
are fixed by fitting very small quantities. In the AV18/UIX 
case with $\Lambda=600$ MeV, for instance, 
the value $\Delta\mu$ is 0.0461(4) for triton,
and --0.0211(4) for $^3$He, where the uncertainties are 
statistical errors due to the Monte Carlo integrations.
Consequently, the resulting values
will be sensitive to several factors, including numerics. However,
it is worthwhile pointing out that the isovector contact contribution
to the muon capture rates under consideration turns out to be 
negligible. 
Finally, we should point out that the present $\chi$EFT model for the
weak vector current operator differs in some of its two-pion exchange
parts from that obtained in time-ordered perturbation theory 
by some of the present authors in Ref.~\cite{Pas09}. The origins of these
differences have been discussed in Refs.~\cite{Pas09,Koe09}. 
However, for consistency with 
the calculations of the $pp$ and $hep$ reactions of Ref.~\cite{Par03}, 
we have chosen to use the $\chi$EFT model illustrated above. 
We do not expect these differences to be numerically significant for
the processes under consideration.
\begin{center}
\begin{table}[tb]
\begin{tabular}{ccccc}
\hline
   & $\Lambda$ (MeV) & $g_{4S}$ & $g_{4V}$ & $d_R$ \\
\hline
   & 500 & 0.69(1) & 2.065(6) & 0.97(7) \\
AV18/UIX 
   & 600 & 0.55(1) & 0.793(6) & 1.75(8) \\
   & 800 & 0.25(2) & --1.07(1) & 3.89(10) \\
\hline
N3LO/N2LO
   & 600 & 0.11(1) & 3.124(6) & 1.00(9) \\
\hline
\end{tabular}
\caption{\label{tab:eft} The LECs $g_{4S}$ and $g_{4V}$ associated with 
the isoscalar and isovector contact terms in the electromagnetic current 
(see Eq.~(\ref{eq:jvctg})), and the LEC $d_R$ 
of the two-body axial-current contact term, calculated for three
values of the cutoff $\Lambda$ with
triton and $^3$He wave functions obtained from the AV18/UIX model.
For $\Lambda=600$ MeV, 
also the N3LO/N2LO model 
is also used.}
\end{table}
\end{center}

\section{Results}
\label{sec:res}
The results for the total rates of muon capture on deuteron 
and $^3$He are presented in the following two subsections.

\subsection{Muon capture on deuteron}
\label{subsec:mud}

In a partial wave expansion
of the final $nn$ state, all channels with total angular momentum $J\leq 2$
and relative orbital angular momentum $L\leq 3$ have been included, i.e.,
$^1S_0$, $^3P_0$, $^3P_1$, $^3P_2$, $^1D_2$ and $^3F_2$. 
Partial waves of higher order contribute less $\simeq 0.5$ \% to the rate.
Indeed, the $^3F_2$ contribution turns out to be
already below this level.

\begin{center}
\begin{table}[tb]
\begin{tabular}{cccccccc}
\hline
SNPA (AV18) & 
$^1S_0$ & $^3P_0$ & $^3P_1$ & $^3P_2$ & $^1D_2$ & $^3F_2$ & Total \\   
\hline
$g_A$=1.2654(42) & 246.6(7) & 20.1 & 46.7 & 71.6 & 4.5 & 0.9 & 390.4(7)\\
$g_A$=1.2695(29) & 246.8(5) & 20.1 & 46.8 & 71.8 & 4.5 & 0.9 & 390.9(7) \\
\hline
EFT* (AV18) &
$^1S_0$ & $^3P_0$ & $^3P_1$ & $^3P_2$ & $^1D_2$ & $^3F_2$ & Total \\   
\hline
$\Lambda=500$ MeV & 250.0(8) & 19.9 & 46.2 & 71.2 & 4.5 & 0.9 & 392.7(8) \\
$\Lambda=600$ MeV & 250.0(8) & 19.8 & 46.3 & 71.1 & 4.5 & 0.9 & 392.6(8) \\
$\Lambda=800$ MeV & 249.7(7) & 19.8 & 46.4 & 71.1 & 4.5 & 0.9 & 392.4(7) \\
\hline
EFT* (N3LO) &
$^1S_0$ & $^3P_0$ & $^3P_1$ & $^3P_2$ & $^1D_2$ & $^3F_2$ & Total \\   
\hline
$\Lambda=600$ MeV & 250.5(7) & 19.9 & 46.4 & 71.5 & 4.4 & 0.9 & 393.6(7) \\
\hline
\end{tabular}
\caption{\label{tab:mudtot} Total rate for muon capture on deuteron, in the
doublet initial hyperfine state, in 
s$^{-1}$. The different partial wave contributions are indicated. The
numbers among parentheses indicate the theoretical uncertainty
arising from the adopted fitting procedures, as explained in 
Sec.~\protect\ref{sec:weak}. Such uncertainty is not indicated when 
less than 0.1 s$^{-1}$. The label ``SNPA'' indicates that the results
have been obtained using the model for the weak transition operator 
of Sec.~\protect\ref{subsec:snpa}, while the label ``EFT*'' indicates
that the model of Sec.~\protect\ref{subsec:eft} is used. 
The AV18 and N3LO interactions have been used to calculate the deuteron 
and $nn$ wave functions.}
\end{table}
\end{center}
We present in Table~\ref{tab:mudtot} the results for the total capture rate
in the doublet hyperfine state.
Both models for the  nuclear weak transition operator presented
in Sec.~\ref{subsec:snpa} and Sec.~\ref{subsec:eft} have been used,
labeled SNPA and EFT*, respectively.
The nuclear wave functions have been calculated with the AV18~\cite{Wir95} 
or the N3LO~\cite{Ent03} two-nucleon interaction. 
The label EFT* is used to denote the results of calculations in which
the matrix elements of $\chi$EFT weak operators are evaluated between
wave functions corresponding to both conventional and chiral potentials.
The first approach (based on a conventional potential) is often referred
to in the literature as the ``hybrid'' approach.  The second, using chiral
potentials and currents, is in principle a full-fledged $\chi$EFT calculation,
except that these potentials and currents have not 
(yet) been derived consistently
at the same order in the low-momentum scale.  For this reason, we characterize
the corresponding results with the EFT* label.

Within each 
Hamiltonian model, the parameters present in the SNPA and EFT* axial current 
models have been fitted to reproduce GT$^{\rm EXP}$ in tritium $\beta$-decay,
as discussed in Sec.~\ref{sec:weak}. 
Furthermore, the LECs in the EFT* weak vector current 
have been fitted to reproduce the $A=3$ magnetic moments.
The three-nucleon wave functions have been generated, 
in this fitting procedure, from 
two- and three-nucleon interactions,
either AV18 and UIX~\cite{Pud95}, or
N3LO and N2LO~\cite{Nav07}. 
Inspection of the table shows that 
the $^1S_0$ contribution is the leading one, but $L \geq 1$
contributions are significant and account for $\sim 37$ \% 
of the total rate. 
By comparison between the first and second row of the table, we conclude that
there is no difference in the results, within uncertainties, 
when the older value for
$g_A$, $g_A$=1.2654(42), or the most recent one, $g_A$=1.2695(29), is used.
This reflects the fact that the factor $R_A$, see Sec.~\ref{subsec:snpa}, 
has been constrained by GT$^{\rm EXP}$ in both cases. 
The model dependence due to interactions, currents, and the
cutoff $\Lambda$ (present in the $\chi$EFT version of these currents)
is at the 1 \% level in the total rate, and hence very weak.
It is a bit larger, 2 \%, in the $^1S_0$ channel, for which 
two-body current contributions are larger, see Tables~\ref{tab:mudsnpa}
and~\ref{tab:mudeft} below.
In conclusion, a total capture rate in the range
\begin{equation}
\Gamma^D = ( 389.7 - 394.3 )\,{\rm s}^{-1} \ ,
\label{eq:totmud}
\end{equation}
can be conservatively ascribed to reaction~(\ref{eq:mud}).
This result is in agreement with the measurements
of Refs.~\cite{Wan65,Ber73,Car86}, but not with that of 
Ref.~\cite{Bar86}. The differences with the theoretical results of 
Refs.~\cite{Tat90,Doi90-91,And02} are also very small. In particular, 
the authors of Ref.~\cite{And02} find $\Gamma^D(^1S_0)=245$ s$^{-1}$, 
in contrast to $\Gamma^D(^1S_0)=250$ s$^{-1}$ reported here. 
We have explicitly verified that this difference is
mainly due to the inclusion in the present work of mesonic two-body 
contributions in the weak vector current 
beyond the soft one-pion exchange term
discussed in Sec.~\ref{subsec:eft}. On the other hand,
the results of Refs.~\cite{Ada90,Ric10} are significantly larger
than those listed here, presumably because these authors
have not constrained their weak current to reproduce GT$^{\rm EXP}$ 
and the isovector magnetic moment of the trinucleons.
We observe that our approach also provides a value for
muon capture on $^3$He in 
excellent agreement with the experimental datum.

\begin{figure*}[tbp]
\begin{center}
\includegraphics[width=\textwidth,height=0.5\textheight]{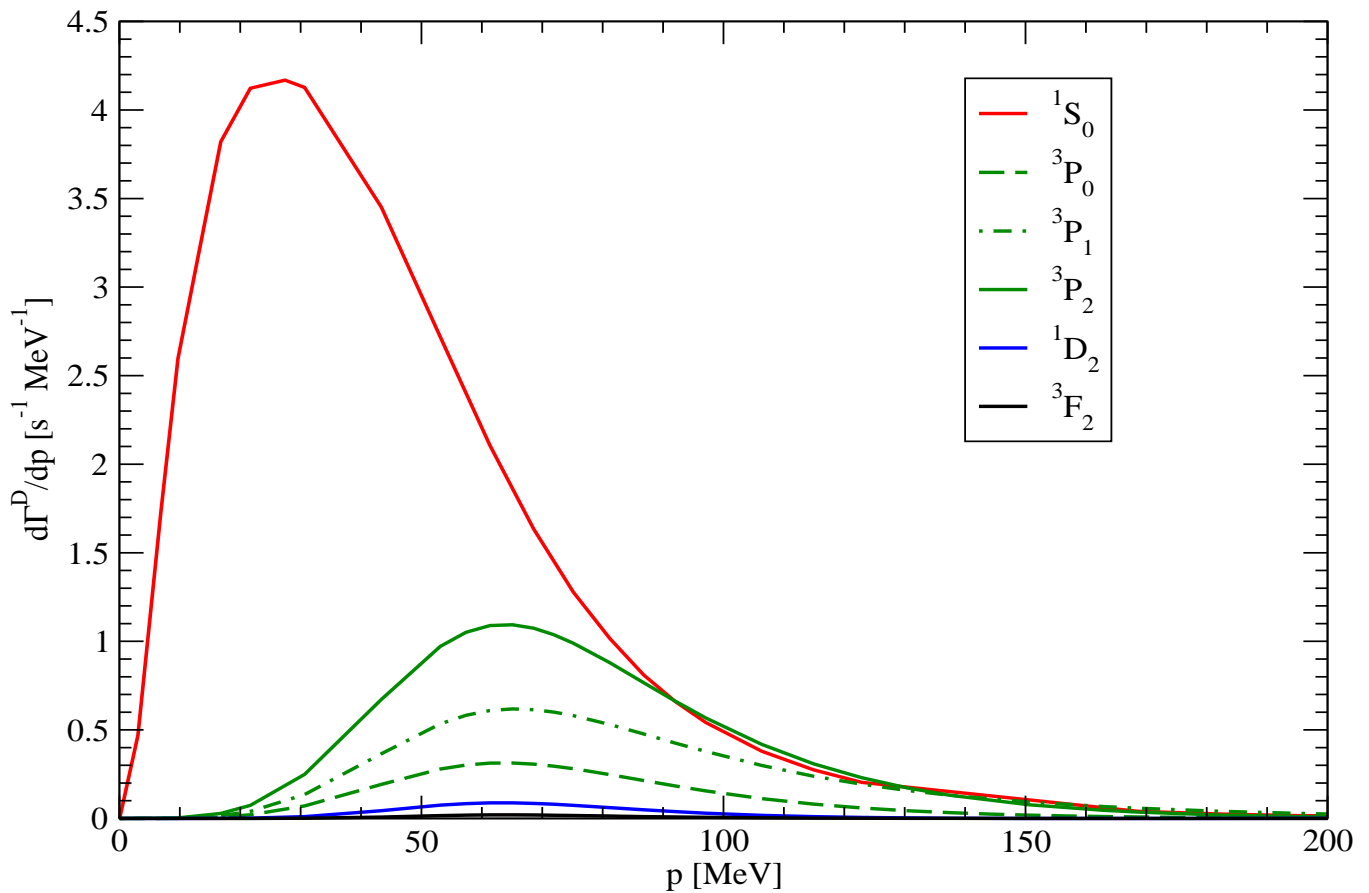}
\end{center}
\vspace*{0.5cm}
\caption{\label{fig:dcap} (Color online)
Differential capture rate $d\Gamma^D/dp$, defined in Eq.~(\ref{eq:dgd}),
as function of the $nn$ relative momentum $p$ in MeV. The calculation
is performed with the SNPA (AV18) model.}
\end{figure*}

For future reference, we show in Fig.~\ref{fig:dcap}
the differential capture rate $d\Gamma^D/dp$, defined in Eq.~(\ref{eq:dgd}),
as function of the $nn$ relative momentum $p$, 
calculated with the SNPA (AV18) model. 
Integrating each partial wave contribution leads to the values
listed in the first row of Table~\ref{tab:mudtot}.

\begin{center}
\begin{table}[tbh]
\begin{tabular}{ccc}
\hline
 & $^1S_0$ & $^3P_2$ \\
\hline
IA+RC & 293.0 & 82.6 \\
PS & 233.2 & 71.6 \\
mesonic & 237.5 & 71.4 \\
$\Delta$-PT($A$-w/o PS)+$\Delta$-PT($V$) & 248.3(8) & 71.6 \\
$\Delta$-PT($A$-w/o PS)+$\Delta$-TCO($V$) & 247.7(8) & 71.6 \\
$\Delta$-PT($A$-w PS)+$\Delta$-TCO($V$) & 246.6(7) & 71.6 \\
\hline
\end{tabular}
\caption{\label{tab:mudsnpa} Cumulative contributions to the 
total rate for muon capture on deuteron in 
s$^{-1}$, when the model of Sec.~\protect\ref{subsec:snpa} is used
for the weak transition operator. The deuteron and $nn$ wave functions
are calculated using the AV18 interaction. Only the $^1S_0$ and $^3P_2$
partial waves are considered, since they give the leading contributions.
The labels ``IA+RC'', ``PS'', and ``mesonic'' indicate the results
obtained by retaining the impulse approximation plus relativistic 
corrections, the induced pseudoscalar, and the purely mesonic contributions.
The label ``$\Delta$-PT($A$-w/o PS)'' is used to indicate that the axial 
charge and current contributions due to $\Delta$-isobars 
are retained perturbatively, but the induced pseudoscalar contribution
in the one-body $\Delta$-current is not included, while the 
label ``$\Delta$-PT($A$-w PS)'' is used when the $N$-to-$\Delta$
induced pseudoscalar
contribution is also included. Finally, the labels ``$\Delta$-PT($V$)'' 
and ``$\Delta$-TCO($V$)'' are used when the vector current 
contributions due to $\Delta$-isobars are treated perturbatively and
within the TCO scheme, respectively. The value $g_A$=1.2654(42) is used.
The numbers in parentheses indicate the theoretical errors arising 
from the fitting procedure adopted for $R_A$. This uncertainty is not shown 
if less than 0.1 s$^{-1}$.}
\end{table}
\end{center}
\begin{center}
\begin{table}[bth]
\begin{tabular}{cccc}
\hline
 & & $^1S_0$ & $^3P_2$ \\
\hline
 & IA+RC & 292.5 & 82.5 \\
$\Lambda=500$ MeV & PS & 232.7 & 71.6 \\
 & mesonic & 250.0(8) & 71.2 \\
\hline
 & IA+RC & 292.5 & 82.5 \\
$\Lambda=600$ MeV & PS & 232.7 & 71.6 \\
 & mesonic & 250.0(8) & 71.1 \\
\hline
 & IA+RC & 292.5 & 82.5 \\
$\Lambda=800$ MeV & PS & 232.7 & 71.6 \\
 & mesonic & 249.7(7) & 71.1 \\
\hline
\end{tabular}
\caption{\label{tab:mudeft} Same as Table~\protect\ref{tab:mudsnpa}
but with the nuclear current model of Sec.~\protect\ref{subsec:eft}.
The labels ``IA+RC'', ``PS'', and ``mesonic'' indicate the results
obtained by retaining the impulse approximation plus relativistic 
corrections, the induced pseudoscalar, and the mesonic contributions.
The relativistic corrections to the one-body vector current operator
include the additional terms of Eq.~(\protect\ref{eq:jv1brc}).
The numbers in parentheses indicate the theoretical errors arising 
from the fitting procedure adopted for 
$d_R$ and $g_{4V}$. This uncertainty is not shown 
if less than 1 s$^{-1}$.}
\end{table}
\end{center}
The cumulative contributions to the leading $^1S_0$ and $^3P_2$ capture rates
of various terms in the nuclear current operator
are given in Table~\ref{tab:mudsnpa} for the SNPA 
and~\ref{tab:mudeft} for the EFT*.
Table~\ref{tab:mudsnpa} shows that:
(i) the (one-body) induced pseudoscalar contribution is significant,
as expected, and reduces the rate of $\sim$ 20 \% 
($\sim$ 13 \%) in $^1S_0$ ($^3P_2$) capture;
(ii) the mesonic contributions are very small; (iii) the $\Delta$-isobar 
contributions are significant only in $S$-wave capture, and
differences in the vector two-body current originating from treating the 
$\Delta$-isobar degrees of freedom either perturbatively or within the TCO 
approach are negligible; (iv) the $\Delta$-isobar 
induced pseudoscalar axial current contribution of 
Eqs.~(\ref{eq:j1ND}) and~(\ref{eq:gpsd})
is of the order of 1 s$^{-1}$ for $^1S_0$ capture.

From inspection of Table~\ref{tab:mudeft}, we can conclude that:
(i) the one-body contributions (labeled ``IA+RC'' and ``PS'') are slightly
different from those reported in the SNPA.
While the EFT* one-body axial charge and current, and vector charge, 
operators are the same as in the SNPA, the EFT* 
one-body vector current includes the relativistic 
corrections of Eq.~(\ref{eq:jv1brc}), ignored
in the SNPA. On the other hand, the differences between
SNPA and EFT* one-body contributions remain well below
the 1 \% level, and therefore within the theoretical uncertainty.
However, it would be interesting to study the contributions of 
these relativistic corrections to the many electromagnetic
observables analyzed within the SNPA over the years~\cite{Mar05,Mar08}.
A first step in this direction has been done in Ref.~\cite{Gir10}.
(ii) The $\Lambda$ dependence is very weak, well below
the 1 \% level. (iii) The $^3P_2$ term does not show any $\Lambda$ dependence
nor any sensitivity to the $d_R$ and $g_{4V}$ fitting procedure. This 
is easily understood by observing that the contact terms in the 
vector and axial
currents are proportional to $({\bm \tau}_i \times {\bm \tau}_j)_- 
({\bm \sigma}_i \times {\bm \sigma}_j)$ (see Ref.~\cite{Par03} 
and Eq.~(\ref{eq:jvctg})), whose matrix elements vanish 
when calculated between the 
deuteron state and $^3P_J$ $nn$ states.

By comparison of the results listed in Tables~\ref{tab:mudsnpa} 
and~\ref{tab:mudeft}, we observe that the main source for the 2 \%
theoretical uncertainty in the $^1S_0$ channel quoted above is the
$\Delta$-isobar induced pseudoscalar contribution: were it to be neglected,
the SNPA and EFT* $^1S_0$ results would agree at the 1 \% level. 
We observe that the strength of the associated operator, which is only
included in the SNPA calculation and plays no role in tritium
$\beta$-decay (being proportional to the momentum transfer ${\bf q}$),
follows from PCAC and pion-dominance arguments, see Sec.~\ref{subsec:snpa}.

\subsection{Muon capture on $^3$He}
\label{subsec:muhe3}

We present in Table~\ref{tab:mu3tot} the results for the total capture 
rate $\Gamma_0$ defined in Eq.~(\ref{eq:g0}).
The theoretical uncertainties due
to the fitting procedure of $R_A$ (SNPA) or $d_R$ and $g_{4V}$ (EFT*)
are in parentheses.
The results in the first two rows 
have been obtained in the SNPA, using the AV18/UIX Hamiltonian and
the two available values for $g_A$. These
results are the same, since a change in $g_A$ is
compensated by a corresponding one in $R_A$. 
The results of the next three rows have been obtained
within the hybrid approach EFT* (AV18/UIX). They
show a very weak $\Lambda$-dependence, and are in excellent agreement
with those reported in SNPA. The result in the last row is with the
EFT* (N3LO/N2LO).
The EFT* (AV18/UIX) and EFT* (N3LO/N2LO) differ by 
8 s$^{-1}$, or less than 1 \%.
In view of this, we quote conservatively a 
total capture rate for reaction~(\ref{eq:mu3})
in the range
\begin{equation}
\Gamma_0=(1471-1497)\,{\rm s}^{-1}\ ,
\label{eq:totmu3}
\end{equation}
by keeping the lowest and upper bounds in the values of 
Table~\ref{tab:mu3tot}.
\begin{center}
\begin{table}[bth]
\begin{tabular}{cc}
\hline
SNPA (AV18/UIX) & $\Gamma_0$ \\   
\hline
$g_A$=1.2654(42) & 1486(8) \\
$g_A$=1.2695(29) & 1486(5) \\
\hline
EFT* (AV18/UIX) & $\Gamma_0$ \\   
\hline
$\Lambda=500$ MeV & 1487(8) \\
$\Lambda=600$ MeV & 1488(9) \\
$\Lambda=800$ MeV & 1488(8) \\
\hline
EFT* (N3LO/N2LO) & $\Gamma_0$ \\   
\hline
$\Lambda=600$ MeV & 1480(9) \\
\hline\hline
EXP. & 1496(4) \\
\hline
\end{tabular}
\caption{\label{tab:mu3tot} Total rate for muon capture on $^3$He, in 
s$^{-1}$. The
numbers in parentheses indicate the theoretical uncertainties
due to the adopted fitting procedure, see
Sec.~\protect\ref{sec:weak}. 
The label ``SNPA'' (``EFT*'') indicates that the results
have been obtained using the model for the weak transition operator 
of Sec.~\protect\ref{subsec:snpa} (Sec.~\protect\ref{subsec:eft}).
The triton and $^3$He 
wave functions are obtained from the AV18/UIX and 
N3LO/N2LO Hamiltonians.}
\end{table}
\end{center}

The contributions of the different components of the weak current and
charge operators to the total rate and to the reduced matrix elements
(RMEs) of the contributing multipoles are reported in 
Table~\ref{tab:mu3snpa} for SNPA
and~\ref{tab:mu3eft} for EFT*.
The HH wave functions have been calculated using the
AV18/UIX Hamiltonian model. 
In Table~\ref{tab:mu3snpa}, $R_A$ has been fixed to its central value
1.21, and in Table~\ref{tab:mu3eft} we use $\Lambda=600$ MeV and 
consequently $d_R=1.75$ and $g_{4V}=0.793$ (see Table~\ref{tab:eft}).
The notation in these tables is the same as that of 
Tables~\ref{tab:mudsnpa} and~\ref{tab:mudeft} of Sec.~\ref{subsec:mud}, 
with the only exception
of the label ``$\Delta$-TCO(V)'', which here indicates that
the two-body $\Delta$-currents are treated with the TCO method, and
that the three-body current, constructed consistently with the UIX
three-nucleon interaction, are also included.
In Table~\ref{tab:mu3snpa}, we observe
that the induced pseudoscalar term gives a significant contribution to 
the $L_1(A)$ and $C_1(A)$ RMEs, although the latter is much smaller
than $L_1(A)$ in magnitude. The mesonic contributions to
$C_0(V)$, $C_1(A)$, $L_1(A)$, and $E_1(A)$ are small, while they provide
a 15 \% correction to $M_1(V)$, as expected (in the isovector 
magnetic moments of the trinucleons, these mesonic currents give 
a 15 \% contribution relative to the one-body). Contributions
due to $\Delta$-isobar excitations in the weak vector and 
axial currents are at the few \% level.
Finally, we note that the result for $\Gamma_0$ is in excellent agreement 
with that of the earlier study of Ref.~\cite{Mar02}. This agreement comes 
about because of two compensating effects: on the one hand, 
the $L_1(A)$ and $E_1(A)$ RMEs are, in magnitude, slightly larger, 
as consequence of the fact that the parameter $R_A$ is slightly larger
here than in Ref.~\cite{Mar02}, because of the more accurate wave functions
employed. On the other hand, the $M_1(V)$ RME corresponding to
$\Delta$-TCO($V$) is slightly smaller than
calculated in Ref.~\cite{Mar02}, $M_1(V)=0.1355$ in Ref.~\cite{Mar02}
versus $M_1(V)=0.1346$. This last value 
results from dividing the value listed in
Table~\ref{tab:mu3snpa} by the factor 1.022, arising from 
the normalization
correction of the trinucleon wave functions due to the presence of
explicit $\Delta$-isobar degrees of freedom, i.e.,
$N_\Delta=\sqrt{\langle \Psi_{N+\Delta} | \Psi_{N+\Delta} \rangle /
\langle \Psi_{N {\rm only}} | \Psi_{N {\rm only}} \rangle}$,
where $\Psi_{N+\Delta}$ ($\Psi_{N {\rm only}}$) 
is the nuclear wave function with both nucleon
and $\Delta$ (nucleon only) degrees of freedom.  
Indeed, the $\Delta$-TCO($V$) $M_1(V)$ RME  
should be compared with $M_1(V)$ corresponding to
$\Delta$-PT($V$) in Table~\ref{tab:mu3snpa}.

In Table~\ref{tab:mu3eft}, we observe that:
(i) the one-body vector current contribution to the $M_1(V)$ RME is
different than in SNPA. This is due to the presence of the additional
relativistic corrections of Eq.~(\ref{eq:jv1brc}).
(ii) The mesonic contributions are significant for the $L_1(A)$,
$E_1(A)$ and $M_1(V)$ RMEs, and bring their values closer to those in the 
SNPA. 
(iii) The $1\pi C$, $2\pi$, and contact terms in the
mesonic vector current are important. If they were to be neglected,
the total $M_1(V)$ RME would be equal to 0.1201 and 
consequently $\Gamma_0=1453$ s$^{-1}$. 
(iv) The results of Table~\ref{tab:mu3eft} should be compared with those
of Table 2 of Ref.~\cite{Gaz08}. We find significant differences
in all the RMEs, both for the one-body contribution (here labeled ``PS''
and in Ref.~\cite{Gaz08} ``IA'') and the complete calculation. Only
the results for the $M_1(V)$ RME appear to be similar to each other, although
it is unclear whether in Ref.~\cite{Gaz08} the vector $1\pi C$, $2\pi$ and 
contact-term contributions are included. 

Finally, we observe that when the N3LO/N2LO Hamiltonian
model is used, $C_0(V)$, $C_1(A)$, $L_1(A)$, $E_1(A)$ and $M_1(V)$
are --0.3288, 0.4130$\times 10^{-2}$, --0.2773, --0.5810 
and 0.1329, respectively, when the mesonic contributions are included.
Comparing these results with the AV18/UIX ones of Table~\ref{tab:mu3eft},
we see that all the RMEs are comparable at the 1 \% level. This fact is
reflected in the $\sim$ 1 \% difference between the AV18/UIX and N3LO/N2LO
results for $\Gamma_0$. Therefore, it does not seem possible to identify a
particular source for this 1 \% difference.

\begin{widetext}
\begin{center}
\begin{table}[bth]
\begin{tabular}{ccccccc}
\hline
 & $\Gamma_0$ & $C_0(V)$ & $C_1(A)$ & $L_1(A)$ & $E_1(A)$ & $M_1(V)$ \\
\hline
IA+RC & 1530 & --0.3287 & 0.7440$\times$10$^{-2}$ & --0.4056 & --0.5516 & 0.1127 \\
PS & 1316  & & 0.3986$\times$10$^{-2}$ & --0.2589 & \\
mesonic 
& 1385 & --0.3283 & 0.4024$\times$10$^{-2}$& --0.2619 & --0.5562 & 0.1315 \\
$\Delta$-PT($A$-w/o PS)+$\Delta$-PT($V$) 
& 1501 & & 0.4287$\times$10$^{-2}$ & --0.2811 & --0.5821 & 0.1370 \\
$\Delta$-PT($A$-w/o PS)+$\Delta$-TCO($V$) 
& 1493 & & & & & 0.1376 \\
$\Delta$-PT($A$-w PS)+$\Delta$-TCO($V$) & 
1486 & & & --0.2742 & & \\
\hline
\end{tabular}
\caption{\label{tab:mu3snpa} 
Cumulative contributions to the total rate $\Gamma_0$ for muon capture 
on $^3$He, in s$^{-1}$, and to the reduced matrix elements (RMEs) 
$C_0(V)$, $C_1(A)$, $L_1(A)$, $E_1(A)$ and $M_1(V)$ 
(see Sec.~\protect\ref{sec:obs}), when the model of 
Sec.~\protect\ref{subsec:snpa} is used
for the weak transition operator ($R_A=1.21$). 
The HH wave functions have been calculated using the AV18/UIX
Hamiltonian. Notation is the same as in 
Table~\protect\ref{tab:mudsnpa}, except for ``$\Delta$-TCO($V$)'', which 
here is
used to indicate that the $\Delta$-isobar contributions
are treated with the TCO method, and, in addition, three-body weak
vector current contributions are included.
Note that $C_0(V)$ is purely real, while the other
RMEs are purely imaginary. }
\end{table}
\end{center}
\end{widetext}

\begin{center}
\begin{table}[bht]
\begin{tabular}{ccccccc}
\hline
 & $\Gamma_0$ & $C_0(V)$ & $C_1(A)$ & $L_1(A)$ & $E_1(A)$ & $M_1(V)$ \\
\hline
IA+RC & 1517 & --0.3287 & 0.7440$\times$10$^{-2}$& --0.4056 & -0.5516 & 0.1082 \\
PS    & 1303 &          & 0.3986$\times$10$^{-2}$& --0.2589 &         &        \\
mesonic & 
        1488 &          & 0.3978$\times$10$^{-2}$& --0.2810 &-0.5833  & 0.1317 \\
\hline
\end{tabular}
\caption{\label{tab:mu3eft} Same as Table~\protect\ref{tab:mu3snpa}
but with the nuclear current model of Sec.~\protect\ref{subsec:eft}.
The value for the cutoff $\Lambda$ has been fixed to 600 MeV,
and $d_R=1.75$ and $g_{4V}=0.793$.
Notation is the same as in Table~\ref{tab:mudeft}.
}
\end{table}
\end{center}

\section{Summary and conclusions}
\label{sec:sum}

Total rates for muon capture on deuteron and $^3$He have been calculated
within a consistent approach, based on 
realistic interactions and weak currents
consisting of vector and axial-vector components with one- and many-body
terms. Two different approaches have been used to derive these operators:
the first one goes beyond the impulse approximation, by including 
meson-exchange current contributions and terms arising
from the excitation of $\Delta$-isobar degrees of freedom.
This approach, labeled SNPA, has been widely and successfully
used in studies of 
electroweak processes (see for instance 
Refs.~\cite{Sch98,Mar00,Mar02,Mar05,Mar09}). The
second approach, labeled EFT*, includes two-body contributions,
beyond the impulse approximation, derived within a systematic $\chi$EFT
expansion, up to N3LO. The only parameter in the SNPA 
nuclear weak current model is present in the axial current and
is determined by fitting
the experimental value for the triton half-life. In the case of the 
EFT* approach, two LECs appear, one in the vector and one in the axial-vector 
component (note that the LEC appearing in the isoscalar
electromagnetic contact term does not contribute here).
They are fixed to reproduce, respectively, 
the $A=3$ isovector magnetic moment and triton half-life. 

Our final results are summarized in Eqs.~(\ref{eq:totmud}) 
and~({\ref{eq:totmu3}). The very 
accurate experimental datum 
of Ref.~\cite{Ack98} for the total rate in muon capture on $^3$He 
is very well reproduced. For the muon capture on deuteron, 
a precise measurement should become available in the near future~\cite{And07}.
The dependence of the results on the input Hamiltonian model, or on
the model for the nuclear transition operator, is weak, at less than
1 \% level. This weak model dependence is a consequence of the procedure
adopted to constrain the weak current.

We conclude by noting that (i) within the EFT* approach, the two-body
vector current operators beyond the one-pion-exchange
term give significant contributions to the total rate, especially for muon
capture on $^3$He.  These terms have been included in the study of weak
processes here for the first time.  However, since they are proportional
to the momentum transfer, their contribution is expected to be negligible
in the $pp$ and $hep$ reactions~\cite{Par03}.  
(ii) Some of the radiative corrections
are accounted for in the present predictions for the muon rates, since
the value adopted for the Fermi coupling constant $G_V$ is that extracted
from an analysis of superallowed $\beta$-decays~\cite{Har90}.  
The size of these corrections
(roughly 2.4 \% on $G_V^2$) is in agreement with that calculated
independently in Ref.~\cite{Cza07} for the muon capture on hydrogen
and helium.  
The additional radiative corrections, originating from vacuum
polarization effects on the muon bound-state wave function, have also been
estimated by the authors of Ref.~\cite{Cza07}.
In the case of the muon capture on $^3$He,
they increase the predicted rates by about 0.68\%, leading to
1496 s$^{-1}$ for the AV18/UIX model, and to 1490 s$^{-1}$ for the
EFT$^*$(N3LO/N2LO) model, and thus bringing them into closer agreement
with experiment.  (iii) The value for the induced pseudoscalar coupling
used in the present study, $g_{PS}=-8.28$ at the four-momentum transfer
relevant for muon capture on $^3$He ($q_\sigma^2=0.954 m_\mu^2$), is
consistent
with that predicted by PCAC and chiral perturbation theory~\cite{chiPT}.
The agreement between the calculated and experimental muon capture rates
confirms the validity of these predictions.  (iv) The agreement 
between our final results for the muon capture rate on $^3$He
and those of Ref.~\cite{Gaz08} confirms the tight limits obtained there
on possible contributions from second class currents.

Finally, we remark that it would be interesting to extend these
calculations to the processes
$\mu^-+\,^3{\rm He}\rightarrow n+d+\nu_\mu$ 
and $\mu^-+\,^4{\rm He}\rightarrow n+\,^3{\rm H}+\nu_\mu$, for 
which experimental data are also available~\cite{Kuh94,muhe4}.

\acknowledgments
One of the authors (R.S.) would like to thank the Physics Department
of the University of Pisa, the INFN Pisa branch, and especially
the Pisa group for the continuing support and warm hospitality,
extended to him over the past several years. 

The work of R.S. is supported by the U.S. Department of Energy, Office
of Nuclear Science, under contract
DE-AC05-06OR23177.

\end{document}